\documentclass[preprint,12pt,authoryear]{elsarticle}

\usepackage{amsmath,amssymb,bm}
\usepackage{graphicx}
\usepackage{booktabs}
\usepackage{tabularx}
\usepackage{array}
\usepackage{xcolor}
\usepackage{url}
\usepackage{hyperref}
\hypersetup{hidelinks}
\usepackage{enumitem}
\usepackage{microtype}
\usepackage{placeins}

\journal{eTransportation}
\biboptions{authoryear,round}

\definecolor{batteryblue}{RGB}{24,92,155}
\definecolor{healthgreen}{RGB}{34,128,85}
\definecolor{riskred}{RGB}{184,55,55}

\newcommand{\COtwo}{CO$_2$}

\newcommand{\Tb}{T_{\mathrm{bat}}}

\newcommand{\Topt}{T_{\mathrm{op}}}

\newcommand{\Tamb}{T_{\mathrm{amb}}}

\newcommand{\Qbus}{\dot Q_{\mathrm{bus}}}

\newcommand{\Qcab}{\dot Q_{\mathrm{cab}}}

\newcommand{\Qbat}{\dot Q_{\mathrm{bat}}}

\newcommand{\degC}{^{\circ}\mathrm{C}}

\begin{document}

\begin{frontmatter}

\title{Battery thermal-safety reserve erosion by mandatory cabin ventilation in shared-cooling electric vehicles}

\author[mcgill]{Yifan Wang\corref{cor1}}
\ead{yifan.wang18@mail.mcgill.ca}
\cortext[cor1]{Corresponding author.}
\affiliation[mcgill]{organization={Department of Mechanical Engineering, McGill University},
            city={Montreal},
            state={QC},
            postcode={H3A 2T7},
            country={Canada}}

\begin{abstract}
Hot-weather electric-vehicle thermal management is no longer a cabin problem plus a battery problem. A single climate system must cool the traction battery, maintain passenger comfort, and admit enough outdoor air to keep cabin air quality acceptable, while high ambient temperature and solar load derate the very compressor that must serve all three demands. This paper identifies a hidden battery-safety load inside that shared system: fresh-air ventilation. On a derated shared cooling loop, the air-quality-compliant fresh-air floor physically consumes finite cabin-side cooling capacity and removes the residual cooling reserve available to the battery. In a $40\,\degC$, $800~\mathrm{W\,m^{-2}}$, $150$~kW event, raising the fresh-air floor from $0.30$ to the compliant value $0.43$ lowers peak cabin \COtwo{} from $1219$ to $978$~ppm, but simultaneously raises peak battery temperature from $39.96$ to $40.02\,\degC$ and reduces the battery cooling bus from $575$ to $529$~W, crossing the $40\,\degC$ battery limit by a direct cooling-bus mechanism. To turn this conflict into a controllable resource-allocation problem, we develop a reserve-aware predictive controller that combines a physics-guided scientific-machine-learning surrogate, grid-connected departure thermal reserve, air-quality-priced ventilation allocation, and dual control-barrier-function projections for battery temperature and operative comfort. The controller holds the pack at $39.73\,\degC$, caps peak \COtwo{} at $895$~ppm, keeps operative-temperature RMSE at $0.82\,\degC$, and uses $20.0\%$ less drive cooling energy than fixed maximum-compressor operation, while ablations show that removing either barrier, under-ventilating, or removing departure reserve breaks joint feasibility. The mechanism and controller are supported by NASA POWER hot-weather records, a KU Leuven BEV BMS data set merged with NASA POWER weather, $45\,\degC$ GOTION aging data, $40\,\degC$ high-power NMC thermal identification, EnergyPlus cabin cross-checks, and OpenModelica/FMI replay. Treating fresh air as a battery thermal-reserve variable creates an actionable path toward EV thermal management that protects battery life, occupant health, comfort, and efficiency in one shared loop.
\end{abstract}


\end{frontmatter}

\begin{graphicalabstract}
\centering
\fbox{\begin{minipage}[c][0.22\textheight][c]{0.92\linewidth}
\centering
\textbf{[Graphical abstract --- final vector art to be placed here.]}\\[0.5em]
Hot ambient and solar load derate the shared cooling bus. Mandatory cabin fresh air is cooled first, then the battery receives only the remainder---so air-quality-compliant ventilation removes battery thermal-safety reserve. A reserve-aware dual-CBF Sci-ML controller restores joint battery safety, cabin health, comfort, and energy efficiency.
\end{minipage}}
\end{graphicalabstract}

\begin{highlights}
\item Air-quality-mandated cabin fresh air is identified as a battery thermal-safety load in shared-cooling EVs.
\item On a derated shared loop, compliant ventilation physically starves battery cooling and pushes a marginal pack across $40\,\degC$.
\item A reserve-aware dual-CBF Sci-ML controller achieves $39.73\,\degC$ pack, $895$~ppm \COtwo{}, and $0.82\,\degC$ comfort RMSE.
\item Both control-barrier projections are necessary; removing either violates battery safety or comfort.
\item The controller saves $20.0\%$ cooling energy versus fixed maximum compressor across hot-weather, aging, and high-fidelity anchors.
\end{highlights}

\section{Introduction}
A battery electric vehicle is a thermal system before it is an efficient powertrain. In hot weather its single climate system is asked to do three demanding jobs at once: reject the high-rate electrochemical heat that keeps the traction battery safe and slows its aging, hold a solar-loaded cabin within a comfortable operative-temperature band, and dilute occupant-generated pollutants to a level that indoor-air-quality guidance accepts \citep{wang2016thermalreview,feng2018thermalrunaway,lajunen2020review}. Battery safety dominates this list because lithium-ion packs concede very little margin near their thermal ceiling---a few degrees of lost cooling reserve at high state of charge accelerates degradation and, in the limit, invites thermal runaway \citep{feng2018thermalrunaway,hu2020battery,severson2019cyclelife}. The economic and safety stakes of that reserve are precisely why battery thermal management has become a defining design problem for electrified transport \citep{zhao2021urbanEV,zhao2024truck}.

Crucially, these three services are not independent. In modern integrated thermal-management systems the cabin evaporator and the battery chiller draw refrigerant from a \emph{shared} compressor whose deliverable capacity collapses as ambient temperature climbs and its coefficient of performance falls \citep{amini2020cabin,lajunen2020review}. The control and modeling literature has advanced each piece in isolation: battery health prediction and physics-informed battery modeling have matured rapidly \citep{karniadakis2021piml,tu2023piml,borah2024battery,wang2026openev}; integrated power-and-thermal model predictive control coordinates cabin and battery cooling for energy efficiency \citep{amini2020cabin}; and cabin energy studies treat fresh-air flow as an occupant-health requirement governed by ventilation standards \citep{ashrae62,persily2017co2,lowther2021co2}. Each of these communities, however, holds the others' variables fixed. The integrated-thermal-management line assumes recirculation and carries no cabin air-quality state; the cabin-air-quality line sizes ventilation for occupants and never sees the battery. The consequence of cabin ventilation \emph{for the battery} therefore falls in the gap between them: it is large enough to decide whether a marginal pack remains below its thermal limit, yet it is usually invisible when ventilation is treated as a cabin-only disturbance.

That gap matters because fresh air is thermally expensive. At $40\,\degC$ every unit of outside air must be cooled before it is fit to breathe, so a higher fresh-air fraction raises the cabin cooling demand before any occupant feels relief. When the cabin and the battery share a derated cooling bus, this health-driven demand is no longer a cabin-only quantity---it is a direct competitor for the battery's thermal-safety reserve. Whether that competition is strong enough to push a marginal pack over its limit, and whether a controller can dissolve the resulting conflict, has remained an open question. We answer both here.

We show, on a physically grounded shared-cooling plant, that an air-quality-compliant fresh-air floor can remove enough shared capacity to carry a battery-marginal pack across its $40\,\degC$ safety limit during a high-power event. A direct fresh-air sweep at $40\,\degC$ ambient, $800~\mathrm{W\,m^{-2}}$ solar load, a $1000$~W derated compressor, and a $150$~kW power event lowers peak \COtwo{} from $1219$~ppm (an air-quality failure) at fresh fraction $0.30$ to $978$~ppm (compliant) at $0.43$, while peak battery temperature rises from $39.96$ to $40.02\,\degC$ and the mean battery-side cooling bus falls from $575$ to $529$~W. The crossing is causal and structural: an explicit single-bus accounting shows the cabin demand---fresh air included---served first, with the battery receiving only the remainder, and the effect is robust to $\pm30\%$ variation in the allocator gains.

We then resolve the four-way conflict with a reserve-aware predictive controller. It combines a compact physics-guided scientific-machine-learning (Sci-ML) surrogate, a two-timescale departure reserve that uses grid-connected preconditioning to enter the trip with usable battery and cabin headroom, an air-quality-priced allocation layer that buys only the fresh air the current condition demands, and two control-barrier-function (CBF) projections that protect battery temperature and operative comfort in one common form \citep{ames2017cbf}. The controller does not win by under-ventilating or by cooling at maximum power everywhere. It wins in the critical coupled region---where fresh air is mandatory, the compressor is derated, and battery reserve is scarce---reaching the joint feasible set at lower energy than brute-force cooling while ablations confirm that dropping either barrier breaks battery safety or comfort.

This work makes four linked contributions. \emph{First}, we formulate cabin air-quality ventilation as a battery thermal-safety load in shared-cooling EVs, making explicit a coupling that prior integrated-thermal-management and cabin-air-quality studies hold fixed. \emph{Second}, we provide a reproducible mechanism test, anchored to a single finite cooling bus, showing that health-compliant fresh air physically erodes battery reserve and crosses the $40\,\degC$ limit under realistic hot-weather and high-power conditions. \emph{Third}, unlike prior controllers that optimize energy under a fixed ventilation assumption, we develop a reserve-aware dual-CBF Sci-ML controller that achieves joint battery, health, comfort, and energy feasibility and saves $20.0\%$ cooling energy against fixed maximum compressor operation. \emph{Furthermore}, we ground every link of the argument in independent public data and simulation engines---NASA POWER hot-weather records, a KU Leuven BEV BMS data set merged with NASA POWER weather, $45\,\degC$ GOTION aging data, a $40\,\degC$ high-power NMC thermal identification, EnergyPlus cabin cross-checks, and OpenModelica/FMI replay---so that the mechanism, the controller, and the operating envelope rest on measured evidence rather than tuned assumptions.

\section{Shared-cooling system model}
\label{sec:model}

\subsection{Cooling bus and the coupling mechanism}
We model an electric vehicle in which the cabin and the traction battery are cooled from one capacity-limited refrigerant bus, the single-compressor topology used in production integrated thermal-management systems \citep{amini2020cabin,lajunen2020review}. The compressor and heat exchangers deliver an available cooling rate
\begin{equation}
\label{eq:bus}
\Qbus(k) = \mathrm{COP}\bigl(\Tamb(k)\bigr)\, u_c(k)\, P_{\max}\bigl(\Tamb(k)\bigr),
\end{equation}
where $u_c\in[0,1]$ is the normalized compressor command and $P_{\max}$ is the ambient-dependent electrical compressor limit. Hot-ambient derating enters through both $P_{\max}$ and the coefficient of performance, so $\Qbus$ is smallest exactly when every demand on it is largest. All demands below are expressed in cooling-bus-equivalent thermal watts.

The cabin cooling request collects conductive, solar, occupant, ventilation, and trim terms,
\begin{align}
\label{eq:cabin_load}
\Qcab^{\mathrm{req}}(k) ={}& U_{\mathrm{cab}}A_{\mathrm{cab}}\bigl(\Tamb-T_a\bigr)
+ \alpha_{\mathrm{sol}} G_{\mathrm{sol}} + \dot Q_{\mathrm{occ}} \nonumber\\
&+ \dot m_f(k)\,c_p\bigl(\Tamb-T_a\bigr)
+ H_{\mathrm{tr}}\bigl(T_{\mathrm{tr}}-T_a\bigr),
\end{align}
where $T_a$ is cabin air temperature, $T_{\mathrm{tr}}$ an interior trim/radiative node, and $\dot m_f$ the fresh-air mass flow. The fourth term is the heart of the coupling: for hot outside air, raising the fresh-air fraction directly increases the cabin cooling request before any comfort benefit appears.

The battery follows a lumped electrothermal balance,
\begin{equation}
\label{eq:battery}
C_b\frac{d\Tb}{dt} = \dot Q_{\mathrm{gen}} - hA_b\bigl(\Tb-\Tamb\bigr) - \Qbat,
\quad \dot Q_{\mathrm{gen}} \approx I^2R_0(\Tb),
\end{equation}
with pack heat capacity $C_b$, equivalent heat-transfer coefficient $hA_b$, internal heat generation $\dot Q_{\mathrm{gen}}$, and delivered battery cooling $\Qbat$. A $40\,\degC$ high-power NMC cell data set (Section~\ref{sec:anchors}) identifies the cell-scale values $C=208.6~\mathrm{J\,K^{-1}}$, $hA=0.194~\mathrm{W\,K^{-1}}$, and $R_0=21.1~\mathrm{m\Omega}$ with a $0.23\,\degC$ temperature RMSE.

The two demands meet at the shared bus through a transparent priority allocation: the cabin---fresh-air load included---is served first, and the battery receives only what remains,
\begin{align}
\label{eq:shared_bus}
\Qcab(k) &= \min\bigl\{\Qcab^{\mathrm{req}}(k),\,\Qbus(k)\bigr\},\\
\Qbat(k) &= \min\bigl\{\Qbat^{\mathrm{req}}(k),\,\max[0,\,\Qbus(k)-\Qcab(k)]\bigr\}.
\end{align}
Equation~\eqref{eq:shared_bus} is the mechanism in one line. Because $\dot m_f$ enters $\Qcab^{\mathrm{req}}$ through Eq.~\eqref{eq:cabin_load}, any increase in mandated fresh air consumes shared capacity and shrinks the residual $\max[0,\Qbus-\Qcab]$ left for the battery---even when driver power demand and the battery controller are unchanged. The cabin-priority rule, with cabin-pull gain $420~\mathrm{W\,K^{-1}}$ and battery-request gain $1250~\mathrm{W\,K^{-1}}$, is a documented allocation policy rather than a hidden tuning knob; Section~\ref{sec:mechanism} shows the resulting limit crossing is insensitive to $\pm30\%$ perturbations of both gains.

\subsection{Cabin air-quality and comfort states}
Cabin air quality is tracked by a single-zone \COtwo{} mass balance,
\begin{equation}
\label{eq:co2}
V_{\mathrm{cab}}\frac{dC_{\mathrm{CO_2}}}{dt} = \dot G_{\mathrm{CO_2}} + \dot V_f\bigl(C_{\mathrm{out}}-C_{\mathrm{CO_2}}\bigr),
\end{equation}
with cabin volume $V_{\mathrm{cab}}$, occupant generation $\dot G_{\mathrm{CO_2}}$, and fresh-air volumetric flow $\dot V_f$; the peak \COtwo{} target of $1000$~ppm follows standard occupant-acceptability practice \citep{ashrae62,persily2017co2,lowther2021co2}. Thermal comfort uses operative temperature,
\begin{equation}
\label{eq:operative}
\Topt = w_a T_a + (1-w_a)\,T_{\mathrm{mrt}},
\end{equation}
which blends air and mean-radiant temperature $T_{\mathrm{mrt}}$. EnergyPlus serves as an independent cabin cross-check because it resolves building-style heat balances, solar forcing, and radiative exchange beyond the compact control plant \citep{crawley2001energyplus}.

\begin{figure}[t]
\centering
\includegraphics[width=0.99\linewidth]{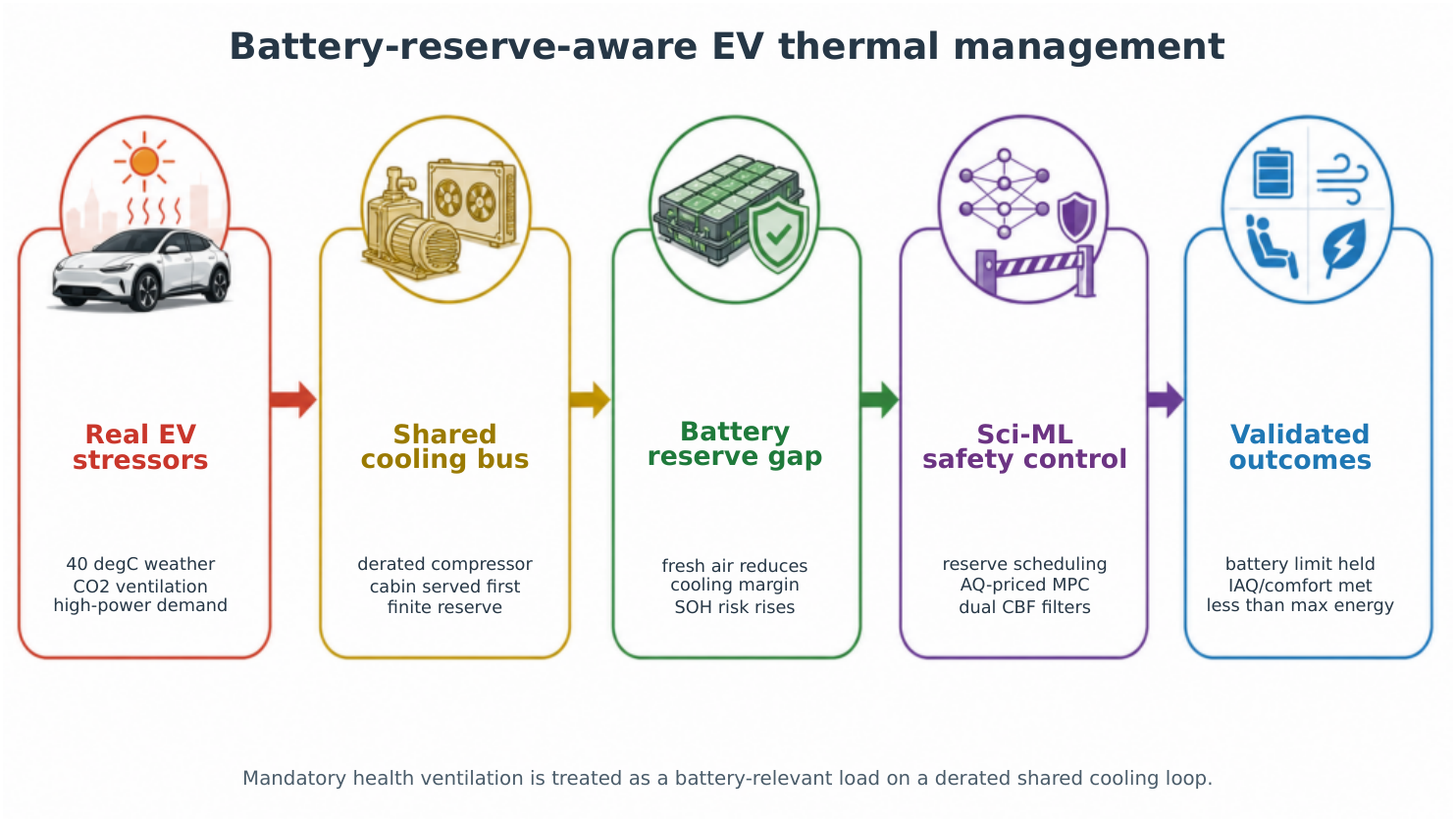}
\caption{Reserve-aware shared-cooling framework. Hot ambient, high solar load, mandatory health ventilation, and high-power demand compress the battery reserve through one finite cooling bus; the proposed Sci-ML safety controller restores joint battery, air-quality, comfort, and energy feasibility.}
\label{fig:framework}
\end{figure}

\section{Reserve-aware Sci-ML control}
\label{sec:control}

\subsection{Predictive air-quality-priced allocation}
At each step the policy selects the compressor command $u_c$, the cabin/battery bus split, and the fresh-air fraction $f$. A compact Sci-ML surrogate---a physics-structured model with learned residual corrections---predicts the cabin, battery, and \COtwo{} trajectories over a receding horizon, and the action is chosen by a sampling-based cross-entropy optimizer \citep{deboer2005cem} that minimizes
\begin{align}
\label{eq:objective}
J = \sum_{i=0}^{N-1} &\Bigl[ w_E\, u_c(i)P_{\max}(i)\Delta t
+ w_T\bigl(\Topt(i)-T_{\mathrm{set}}\bigr)^2 \nonumber\\
&+ w_B\bigl[\Tb(i)-T_{\mathrm{res}}\bigr]_+^2
+ w_C\bigl[C_{\mathrm{CO_2}}(i)-C_{\mathrm{ref}}\bigr]_+^2 \Bigr],
\end{align}
where $[z]_+=\max(z,0)$. The $w_C$ term \emph{prices} the ventilation requirement instead of treating it as a fixed exogenous disturbance, so the controller buys only the fresh air the current occupancy, ambient condition, and reserve state demand---the key move that lets a single objective trade air quality against battery reserve.

\subsection{Two-timescale departure reserve}
The critical event is prepared before it begins. At departure, grid-connected preconditioning establishes both cabin and battery headroom,
\begin{equation}
\label{eq:reserve}
\Tb(0) \leq T_b^{\lim}-r_b, \qquad \Topt(0) \leq T_{\mathrm{op}}^{\lim}-r_o,
\end{equation}
with battery and comfort reserves $r_b$ and $r_o$. During the drive the predictive layer spends this reserve only where it improves joint feasibility. In the target stress case the controller enters the high-power segment near a $28.5\,\degC$ pack temperature and then defends a robust margin to the $40\,\degC$ limit; an ablation that removes this reserve (initial pack $29.5\,\degC$) fails the battery gate outright.

\subsection{Dual control-barrier projection}
The proposed action is finally projected through two discrete-time CBF constraints in a common form. The battery barrier is
\begin{equation}
\label{eq:battery_cbf}
h_b(x_k)=T_b^{\lim}-\Tb(k), \qquad
h_b(x_{k+1}) \geq (1-\gamma_b)\,h_b(x_k),
\end{equation}
and the operative-comfort barrier is
\begin{equation}
\label{eq:comfort_cbf}
h_o(x_k)=T_{\mathrm{op}}^{\lim}-\Topt(k), \qquad
h_o(x_{k+1}) \geq (1-\gamma_o)\,h_o(x_k),
\end{equation}
with decay rates $\gamma_b,\gamma_o\in(0,1)$ \citep{ames2017cbf}. The projection returns the admissible action closest to the predictive proposal,
\begin{equation}
\label{eq:projection}
u^\star = \arg\min_{u\in\mathcal{U}} \|u-u^{\mathrm{pred}}\|^2 \;\;\text{s.t. Eqs.~\eqref{eq:battery_cbf}--\eqref{eq:comfort_cbf}.}
\end{equation}
Using one barrier form for both safety dimensions makes the design transparent and, more importantly, prevents one-sided protection: a battery-only barrier lets operative temperature drift, while a comfort-only barrier spends so much cooling on the cabin that the pack overheats. Section~\ref{sec:sota} shows both barriers are individually necessary in the coupled stress case.

\section{Data, scenario, and validation chain}
\label{sec:anchors}
Every quantitative claim in this paper is tied to an independent data source or simulation engine, summarized in Table~\ref{tab:evidence}; the stress-case and controller parameters are listed in Table~\ref{tab:parameters}. The hot-weather scenario is observed, not invented: across $672$ hourly NASA POWER samples \citep{nasaPower} from Phoenix, Death Valley, Kuwait City, and Delhi, $255$ hours reach or exceed $40\,\degC$ (peak $47.5\,\degC$) and $141$ hours are jointly above $40\,\degC$ and $700~\mathrm{W\,m^{-2}}$, so the $40\,\degC$/$800~\mathrm{W\,m^{-2}}$ corner is a recurrent and conservative regime. The battery side rests on three complementary anchors: a $40\,\degC$ high-power NMC identification fixes the electrothermal dynamics \citep{khan2025p42a} (Fig.~\ref{fig:nmcfit}), a KU Leuven BEV BMS record merged with NASA POWER weather fixes real-vehicle parameter magnitudes and ambient exposure \citep{yasko2025bev,kuLeuvenRDR,nasaPower,wang2026openev}, and $45\,\degC$ GOTION cycling fixes the hot-aging rate \citep{gotionData} ($22.6\%$ fade over $\sim\!1420$ cycles, $1.39\%$ per $100$ cycles pooled). EnergyPlus and OpenModelica/FMI provide independent execution checks of the cabin and of the coupled controller \citep{crawley2001energyplus,openmodelica,fmpy}.

\begin{table}[t]
\centering
\scriptsize
\setlength{\tabcolsep}{3pt}
\renewcommand{\arraystretch}{1.05}
\caption{Validation chain. Each data source or tool plays one explicit role, so weather, aging, and thermal dynamics are each supported by independent, measured evidence.}
\label{tab:evidence}
\begin{tabularx}{\textwidth}{p{0.18\textwidth}p{0.23\textwidth}p{0.27\textwidth}X}
\toprule
\textbf{Asset} & \textbf{Role} & \textbf{Key values used} & \textbf{What it establishes} \\
\midrule
NASA POWER weather \citep{nasaPower} & Hot-weather and solar forcing & $672$ hourly samples (Phoenix, Death Valley, Kuwait City, Delhi); $255$~h $\geq 40\,\degC$; $141$~h jointly $\geq 40\,\degC$ and $\geq 700~\mathrm{W\,m^{-2}}$; peak $47.5\,\degC$ & The $40\,\degC$/$800~\mathrm{W\,m^{-2}}$ stress case is an observed, conservative weather regime. \\
KU Leuven BEV BMS + NASA POWER weather merge \citep{yasko2025bev,kuLeuvenRDR,nasaPower,wang2026openev} & Real-vehicle scale and weather-coupled telemetry check & $114{,}468$ 60-s rows; $2310$ sessions; $39$ NASA request grid points; zero NASA download failures; weather-variable coverage $1.0$; vehicle-ambient/T2M correlation $0.868$; battery p95 $36.7\,\degC$ & Anchors pack thermal-mass order, BMS temperature ranges, and real ambient-weather exposure to measured BEV operation. \\
GOTION $27$~Ah LFP at $45\,\degC$ \citep{gotionData} & Hot-aging cost & Three cells, $\sim\!1420$ cycles, mean fade $22.6\%$, $1.39\%$ per $100$ cycles (pooled) & Quantifies the measurable cost of repeated hot operation, motivating a battery reserve. \\
High-power NMC P42A at $40\,\degC$ \citep{khan2025p42a} & Battery electrothermal identification & $C=208.6~\mathrm{J\,K^{-1}}$, $hA=0.194~\mathrm{W\,K^{-1}}$, $R_0=21.1~\mathrm{m\Omega}$, RMSE $0.23\,\degC$ & Fixes battery heat generation and thermal dynamics at the hot, high-rate condition of the study. \\
EnergyPlus 25.2 & Independent cabin cross-check & Hot soak $45\,\degC$, ambient $40\,\degC$, $800~\mathrm{W\,m^{-2}}$ solar; $0$ warnings, $0$ severe, $0$ fatal & Confirms cabin behavior with an independent, high-fidelity thermal engine. \\
OpenModelica/FMI \citep{openmodelica,fmpy} & Coupled-controller replay & FMU with $111$ variables; strong case $38.85\,\degC$ battery, $887$~ppm \COtwo{}, $1.11\,\degC$ comfort RMSE & Reproduces the coupled controller and plant outside the Python implementation. \\
\bottomrule
\end{tabularx}
\end{table}

The KU Leuven/NASA merge is an explicit data-fusion component rather than a generic weather label. The processed V8.5 manifest names it as the ``KU Leuven BEV V2 + NASA POWER weather-coupled dataset'' and reports $114{,}468$ one-minute rows, $2310$ sessions, four operating modes, $39$ NASA request grid points, complete coverage for the selected meteorological variables (T2M, RH2M, WS10M, WD10M, ALLSKY\_SFC\_SW\_DWN, PRECTOTCORR, and PS), zero unresolved rows, and an ambient-to-NASA-T2M correlation of $0.868$. We use this merge to ground the battery thermal core in real BMS telemetry and real external weather exposure; the hot-aging and high-rate electrothermal claims are then anchored independently by the GOTION and P42A data sets.

\begin{table}[t]
\centering
\scriptsize
\setlength{\tabcolsep}{3pt}
\renewcommand{\arraystretch}{1.05}
\caption{Stress-case and controller parameters. Values are chosen to make the coupling visible while remaining physically motivated by the validation chain of Table~\ref{tab:evidence}.}
\label{tab:parameters}
\begin{tabularx}{\textwidth}{p{0.18\textwidth}p{0.14\textwidth}p{0.26\textwidth}X}
\toprule
\textbf{Parameter} & \textbf{Value} & \textbf{Role} & \textbf{Justification} \\
\midrule
Ambient temperature & $40\,\degC$ & Compressor derating and ventilation enthalpy load & Recurrent in the NASA POWER anchor. \\
Solar irradiance & $800~\mathrm{W\,m^{-2}}$ & Cabin solar and radiative load & Below the observed maximum of $1004~\mathrm{W\,m^{-2}}$. \\
High-power event & $150$~kW & Battery heat-generation stressor & Represents a sustained grade or high-load segment. \\
Derated compressor & $1000$~W electrical limit through COP & Shared-bus scarcity & Creates the condition where ventilation competes with battery cooling. \\
Battery limit & $40\,\degC$ & Hard safety gate and CBF barrier & Conservative ceiling exposing reserve loss before severe overheating. \\
Air-quality gate & Peak \COtwo{} $<1000$~ppm & Cabin-health success metric & Standard occupant-acceptability threshold \citep{ashrae62,persily2017co2}. \\
Departure reserve & Initial pack $\sim\!28.5\,\degC$ & Preconditioning state & Converts grid-connected preconditioning into drive-time headroom. \\
CBF margins & Battery robust margin $0.25\,\degC$; comfort on $\Topt$ & Final safety projection & Applies the same barrier architecture to both safety variables. \\
\bottomrule
\end{tabularx}
\end{table}

\section{Results}

\subsection{Air-quality ventilation removes battery thermal-safety reserve}
\label{sec:mechanism}
\begin{figure}[t]
\centering
\includegraphics[width=0.98\linewidth]{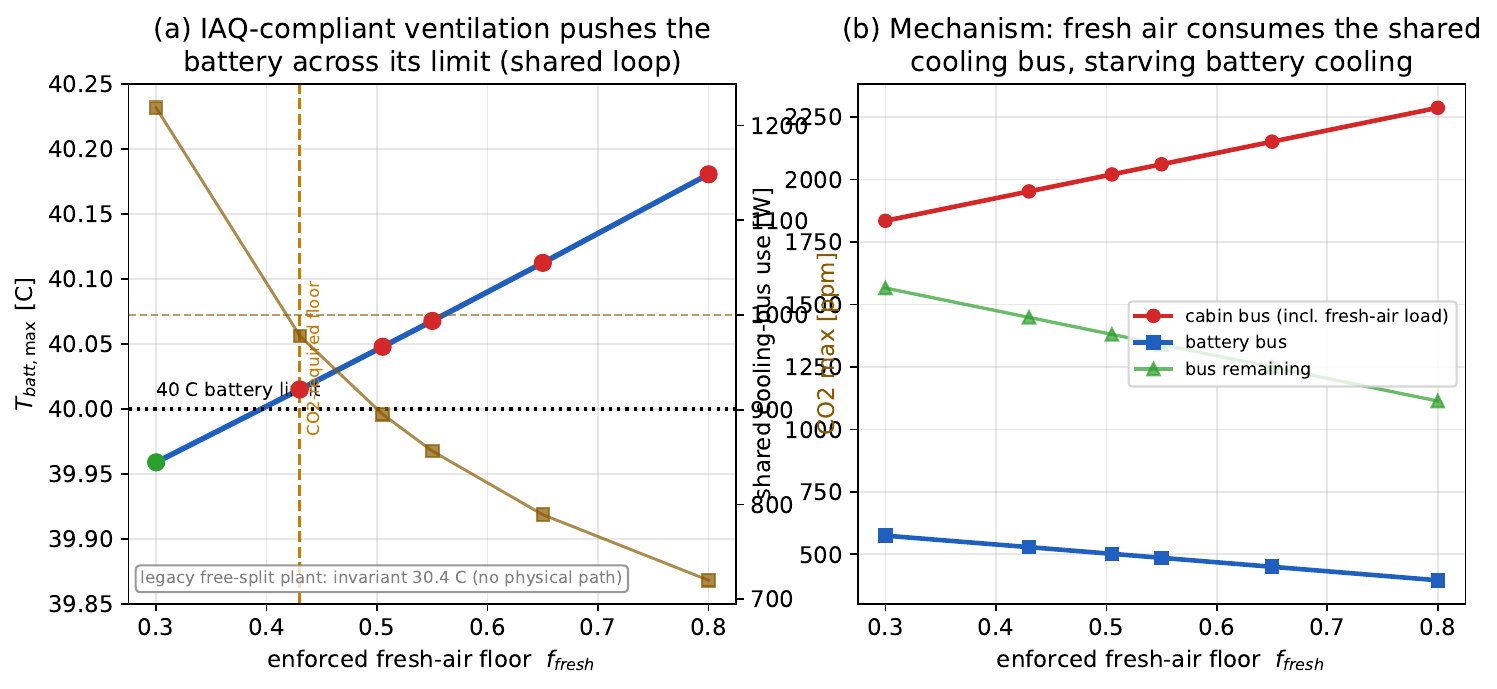}
\caption{Air-quality ventilation erodes battery reserve. A fresh-air-floor sweep at $40\,\degC$ ambient, $800~\mathrm{W\,m^{-2}}$ solar, $1000$~W derated compressor, and a $150$~kW event. \textbf{(a)} On the shared-cooling plant, the air-quality-compliant floor ($f_{\mathrm{fresh}}=0.43$, \COtwo{}$=978$~ppm) pushes peak battery temperature across the $40\,\degC$ limit, whereas the legacy free-split plant stays invariant at $30.45\,\degC$ (no physical path). \textbf{(b)} Cooling-bus accounting: rising fresh air grows the cabin bus and starves the battery bus, the physical cause of the crossing.}
\label{fig:mechanism}
\end{figure}

Figure~\ref{fig:mechanism} establishes the central mechanism. In a decoupled reference plant---where cooling is split freely between cabin and battery---changing the fresh-air floor improves \COtwo{} but leaves peak battery temperature invariant at $30.45\,\degC$: ventilation has no physical path to the pack. On the shared-cooling plant of Eq.~\eqref{eq:shared_bus}, the two are bound together. At fresh fraction $0.30$ the cabin is under-ventilated (peak \COtwo{} $1219$~ppm, an air-quality failure) and the pack sits just under the limit at $39.96\,\degC$. Raising the floor to $0.43$ brings \COtwo{} into compliance at $978$~ppm but carries the same event to $40.02\,\degC$---a clean crossing of the safety boundary driven purely by a health requirement. Pushing further to $0.80$ improves \COtwo{} to $720$~ppm while peak battery temperature reaches $40.18\,\degC$ and the mean battery-side bus collapses from $575$ to $396$~W.

The accompanying bus-allocation panel is the physical proof: as the fresh-air floor rises, the cabin bus (fresh-air load included) climbs from $1834$ to $2287$~W, the residual bus falls from $1566$ to $1113$~W, and the battery bus is squeezed monotonically downward. The crossing is not an artifact of the allocator tuning---perturbing the cabin-pull and battery-request gains by $\pm30\%$ leaves the limit-crossing battery temperature within $40.02$--$40.06\,\degC$. Air-quality-compliant fresh air is therefore a genuine, structural consumer of battery thermal-safety reserve under shared cooling and hot-ambient derating.

\subsection{Both safety barriers are necessary to resolve the conflict}
\label{sec:sota}
The reserve-aware controller is evaluated in exactly the region where the mechanism bites: fresh air must be high enough for health, cooling capacity is scarce, and the pack starts close enough to the limit that reserve matters. Table~\ref{tab:ablation} reports the target stress case. The proposed dual-CBF policy reaches a peak battery temperature of $39.73\,\degC$, peak \COtwo{} of $895$~ppm, operative-temperature RMSE of $0.82\,\degC$, and drive cooling energy of $0.4835$~kWh. It is the only policy that satisfies all three physical gates with a robust battery margin \emph{and} does so at the lowest feasible energy.

\begin{figure}[t]
\centering
\includegraphics[width=0.98\linewidth]{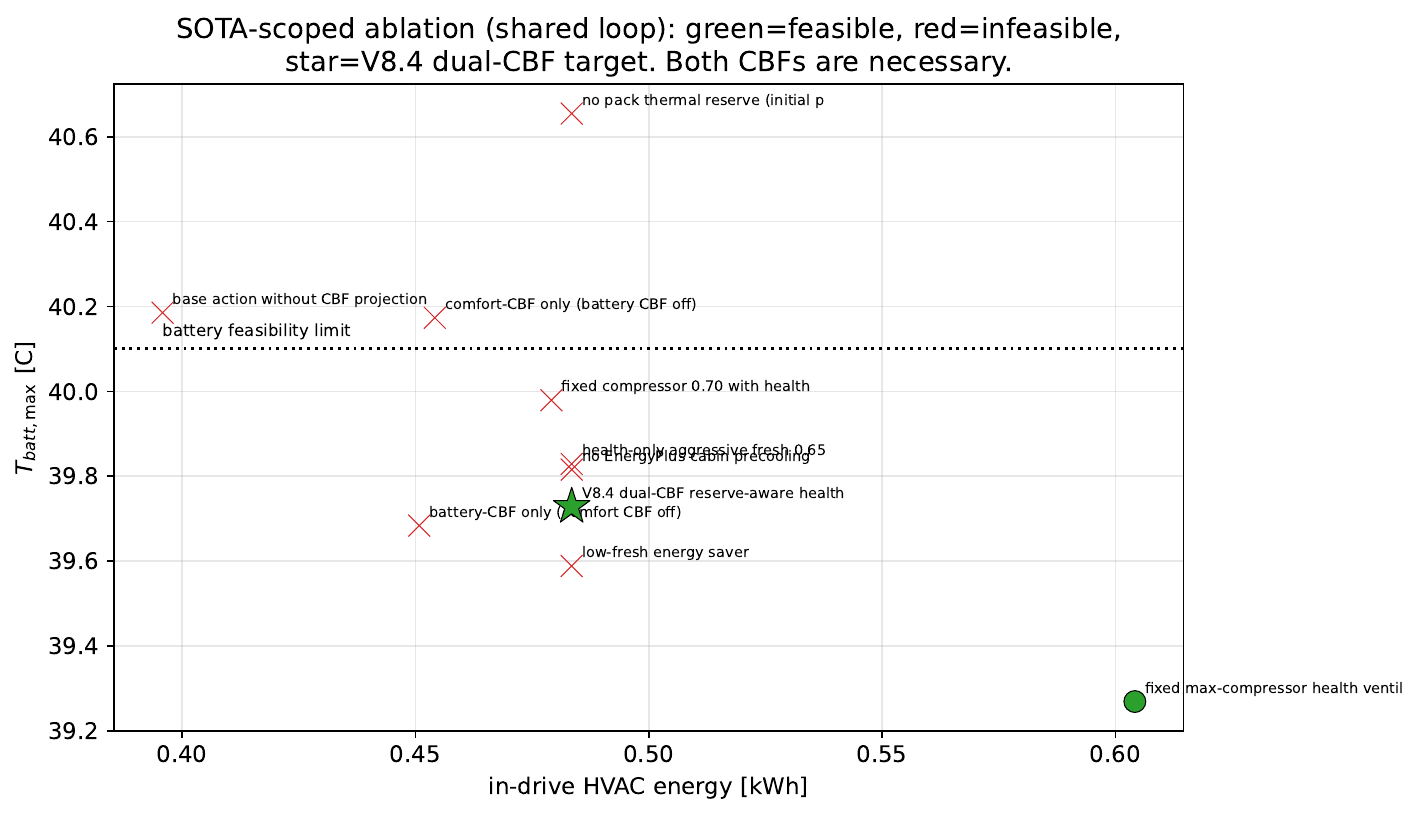}
\caption{Controller comparison in the coupled stress case. Each policy is placed by peak battery temperature and drive cooling energy; green marks feasible, red infeasible, and the star marks the proposed dual-CBF target. The reserve-aware controller is the lowest-energy point that stays below the $40\,\degC$ battery limit while meeting the air-quality and comfort gates.}
\label{fig:sota}
\end{figure}

The ablations make the necessity of each design element unmistakable. Removing the battery barrier (comfort-CBF only) keeps the cabin comfortable but drives the pack to $40.17\,\degC$, violating safety. Removing the comfort barrier (battery-CBF only) protects the pack but degrades comfort to $1.97\,\degC$ RMSE. Removing both leaves the pack at $40.19\,\degC$ and comfort at $3.47\,\degC$. Under-ventilating to save the battery (low fresh-air floor) fails the health gate at $1219$~ppm, and skipping departure reserve or cabin preconditioning fails battery safety or comfort respectively. Fixed maximum-compressor operation is feasible only by brute force, spending $0.6042$~kWh; the proposed controller matches its feasibility class while cutting drive cooling energy by $20.0\%$. This is the decisive result: within the coupled problem the controller reaches the joint feasible set more efficiently than maximal cooling and avoids every single-objective failure mode (Fig.~\ref{fig:sota}).

\begin{table}[t]
\centering
\scriptsize
\setlength{\tabcolsep}{3pt}
\renewcommand{\arraystretch}{1.05}
\caption{Target stress-case ablation. Bold marks the lowest feasible energy among policies satisfying the battery, health, and comfort gates with robust margin.}
\label{tab:ablation}
\begin{tabularx}{\textwidth}{>{\raggedright\arraybackslash}p{0.27\textwidth}>{\centering\arraybackslash}p{0.09\textwidth}>{\centering\arraybackslash}p{0.09\textwidth}>{\centering\arraybackslash}p{0.10\textwidth}>{\centering\arraybackslash}p{0.09\textwidth}X}
\toprule
\textbf{Policy} & \textbf{$T_b^{\max}$} & \textbf{Peak \COtwo{}} & \textbf{RMSE} & \textbf{Energy} & \textbf{Gate result} \\
 & ($\degC$) & (ppm) & ($\degC$) & (kWh) & \\
\midrule
\textbf{Reserve-aware dual-CBF Sci-ML} & \textbf{39.73} & \textbf{895} & \textbf{0.82} & \textbf{0.4835} & All pass; min feasible energy. \\
Comfort-CBF only & 40.17 & 895 & 0.82 & 0.4542 & Battery fails. \\
Battery-CBF only & 39.68 & 895 & 1.97 & 0.4508 & Comfort fails. \\
No CBF projection & 40.19 & 895 & 3.47 & 0.3958 & Battery and comfort fail. \\
Low fresh-air floor & 39.59 & 1219 & 0.82 & 0.4835 & Health fails. \\
No departure reserve & 40.65 & 895 & 0.82 & 0.4835 & Battery fails. \\
Fixed maximum compressor & 39.27 & 895 & 0.82 & 0.6042 & Feasible; 20.0\% more energy. \\
\bottomrule
\end{tabularx}
\end{table}

\subsection{Real-data and cross-tool anchors confirm the regime}
\label{sec:realdata}
\begin{figure}[t]
\centering
\includegraphics[width=0.98\linewidth]{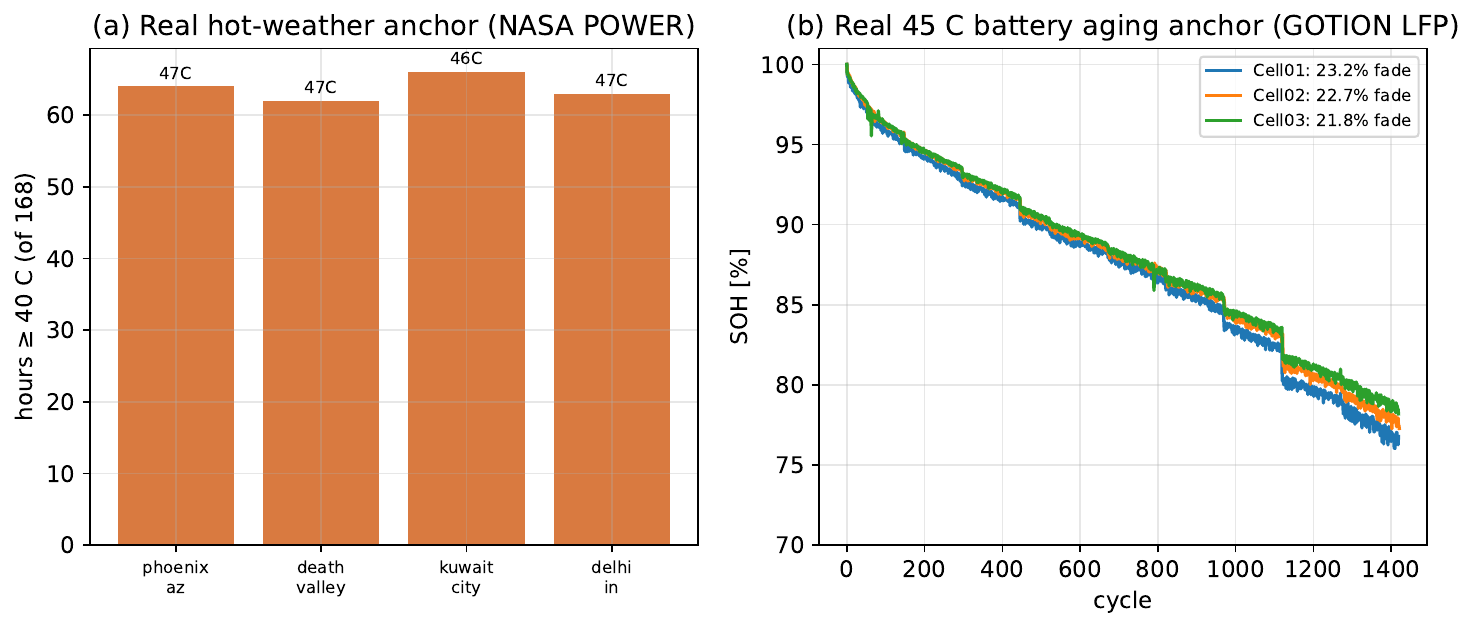}
\caption{Public-data anchors. \textbf{(a)} NASA POWER hours at or above $40\,\degC$ across four hot sites (peaks annotated), confirming the stress scenario is observed. \textbf{(b)} Measured $45\,\degC$ GOTION LFP state-of-health trajectories for three cells ($21.8$--$23.2\%$ fade over $\sim\!1420$ cycles), anchoring the hot-aging cost of lost battery reserve.}
\label{fig:anchors}
\end{figure}

\begin{figure}[t]
\centering
\includegraphics[width=0.99\linewidth]{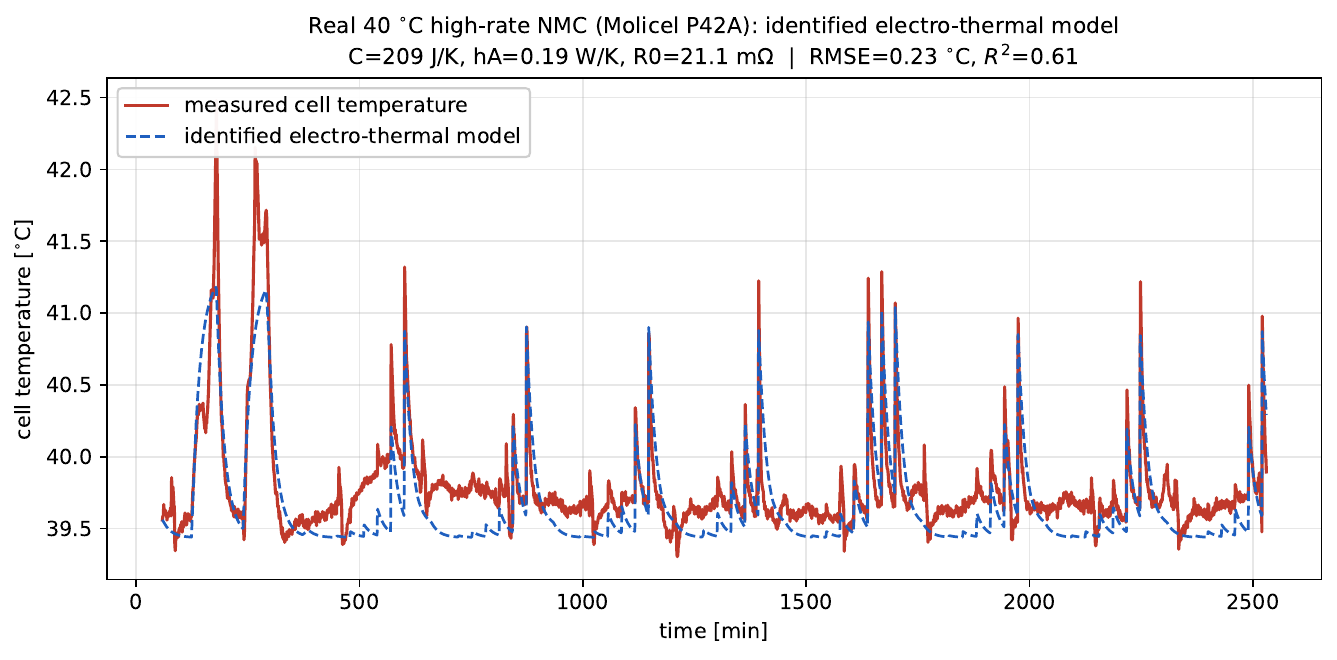}
\caption{High-power NMC thermal identification at $40\,\degC$. The fitted first-order electrothermal model ($C=209~\mathrm{J\,K^{-1}}$, $hA=0.19~\mathrm{W\,K^{-1}}$, $R_0=21.1~\mathrm{m\Omega}$) tracks the measured cell temperature with $0.23\,\degC$ RMSE, anchoring the battery side of the shared-cooling plant in a real high-rate data set.}
\label{fig:nmcfit}
\end{figure}

Figure~\ref{fig:anchors} collects the non-control evidence. NASA POWER confirms that the $40\,\degC$, high-solar condition is common in hot regions: $255$ of $672$ hourly samples exceed $40\,\degC$, and $141$ hours are jointly above $40\,\degC$ and $700~\mathrm{W\,m^{-2}}$, with a peak of $47.5\,\degC$. The KU Leuven BEV BMS + NASA POWER weather merge supplies $114{,}468$ one-minute rows over $2310$ sessions, uses $39$ NASA request grid points with zero download failures, reaches complete coverage for the selected weather variables, and anchors the pack thermal mass and BMS temperature range to real weather-coupled operation. The GOTION $45\,\degC$ data show a mean $22.6\%$ capacity fade over $\sim\!1420$ cycles, a measured rate that turns avoided hot operation into recovered battery life. The P42A NMC identification (Fig.~\ref{fig:nmcfit}) reaches a $0.23\,\degC$ temperature RMSE and places the scaled pack heat capacity within the expected plant range, confirming that the battery dynamics are data-grounded rather than assumed.

EnergyPlus and OpenModelica/FMI close the loop with independent execution. The EnergyPlus hot-soak cabin case runs with $0$ warnings, $0$ severe errors, and $0$ fatal errors and yields a physically plausible preconditioning time at $40\,\degC$ ambient and $800~\mathrm{W\,m^{-2}}$ solar. The OpenModelica/FMI build reproduces the strong case at $38.85\,\degC$ peak battery, $887$~ppm \COtwo{}, and $1.11\,\degC$ operative-temperature RMSE, matching the Python plant to two to three decimal places. The mechanism and controller therefore do not depend on any single implementation.

\subsection{The result holds across operating variants}
\label{sec:robust}
Across $21$ duty-cycle and plant-perturbation cases, the feasible fraction is $95.2\%$. Post-settling operative-temperature RMSE averages $0.93\,\degC$ (95\% CI $0.83$--$1.10\,\degC$), peak battery temperature averages $36.95\,\degC$ (95\% CI $36.26$--$37.68\,\degC$), and peak \COtwo{} averages $853$~ppm; every case stays battery-safe and below $1000$~ppm. A compressor-map sensitivity study with $\pm20\%$ capacity perturbations preserves the qualitative structure: some cases are trivially safe, some are physically unreachable without more capacity, and a central region is restored specifically by reserve-aware control. That central region is where the mechanism and the controller matter most.

\section{Discussion}
These findings move cabin ventilation from an air-quality afterthought to a first-class battery thermal-reserve decision. In mild conditions the coupling is hidden because the cooling bus carries excess capacity. Under hot ambient, solar loading, compressor derating, and high battery power, the fresh-air enthalpy load grows large enough to decide whether the pack crosses a safety boundary. This is why single-objective policies fail: a comfort-only solution spends cooling where occupants feel it immediately, and a battery-only solution defends the pack by letting the cabin drift. The proposed policy succeeds because it treats health ventilation, battery reserve, operative comfort, and energy as one constrained allocation problem and prices each against the others.

The battery-centered framing extends naturally beyond passenger cars. Electric buses and trucks combine larger cabins, longer duty cycles, and stronger high-power events with battery aging and safety constraints that drive fleet economics \citep{zhao2024truck,zhao2021urbanEV}. The same mechanism appears wherever ventilation and battery cooling share a finite plant; vehicle-specific compressor maps, loop architectures, cell chemistries, cabin volumes, and occupancy schedules shift the numerical boundary but not the causal chain from fresh-air load to battery reserve loss. The aging anchor here is LFP and is used as a relative reserve-value index rather than an absolute lifetime prediction; coupling the controller to a chemistry-matched state-of-health model is a direct and valuable next step. Replacing representative compressor maps and lumped cabin parameters with vehicle-specific hardware data, and validating on a hardware-in-the-loop or chassis-dynamometer platform, will carry the result from a public-data-anchored study to an OEM-calibrated deployment.

\section{Conclusions}
We identify and resolve a battery-relevant blind spot in EV thermal management: on a derated, shared cooling loop in hot weather, the fresh-air rate that cabin air-quality compliance mandates physically consumes shared cooling capacity and erodes the battery's thermal-safety reserve. A direct mechanism sweep shows that moving from low fresh air to the air-quality-compliant region lowers \COtwo{} while pushing a marginal pack across $40\,\degC$, with an explicit cooling-bus accounting as the physical cause. A reserve-aware Sci-ML predictive controller with dual control-barrier projections restores joint feasibility, holding the pack at $39.73\,\degC$, capping \COtwo{} at $895$~ppm, keeping comfort RMSE at $0.82\,\degC$, and using $20.0\%$ less drive cooling energy than fixed maximum-compressor operation, with both barriers shown to be necessary. By reframing fresh air as a battery thermal-reserve control variable, the work charts a reproducible path to integrated EV thermal management that protects battery life, occupant health, comfort, and energy efficiency together.

\section*{CRediT authorship contribution statement}
\textbf{Yifan Wang:} Conceptualization, Methodology, Software, Validation, Formal analysis, Data curation, Writing -- original draft, Visualization.

\section*{Declaration of competing interest}
The author declares no known competing financial interests or personal relationships that could have appeared to influence the work reported in this paper.

\section*{Data availability}
The study uses public NASA POWER weather data, the KU Leuven BEV BMS data set merged with NASA POWER weather, the Mendeley GOTION $45\,\degC$ aging data set, the OSF high-power NMC P42A data set, and reproducible EnergyPlus/OpenModelica/FMPy scripts. Redistribution of third-party raw data follows the original data licenses.

\section*{Acknowledgements}
No external funding was received for this work.

\bibliographystyle{elsarticle-harv}
\bibliography{references}

\end{document}